\documentclass{an}
\usepackage{graphicx}
\usepackage{times}
\usepackage{fancyhdr}
\def\gtsim{\raise 2pt \hbox {$>$} \kern-1.1em \lower 4pt \hbox {$\sim$}}
\def\ltsim{\raise 2pt \hbox {$<$} \kern-1.1em \lower 4pt \hbox {$\sim$}}

\begin{document}

\title{Constraining B in galaxy clusters from statistics of giant radio halos}

\author{R. Cassano, G. Brunetti, G. Setti}
\institute{}

\date{Received; accepted; published online}

\abstract{There are several possibilities to constrain the value of the magnetic
field in the ICM, the most direct ones being the combination of inverse Compton and synchrotron observations, and the Faraday rotation measures.
Here we discuss on the possibility to provide constraints on the magnetic field in the ICM 
from the analysis of the statistical properties of the giant radio halos,
Mpc--scale diffuse radio emission 
in galaxy clusters.
Present observations of a few well studied radio halos can be interpreted 
under the hypothesis that the emitting relativistic electrons 
are re-accelerated on their way out. 
By using statistical calculations carried out 
in the framework of the re-acceleration model
we show that the observed radio--power vs cluster mass  
correlation in radio halos can be reproduced only by assuming 
$\mu$G fields in the ICM and a
scaling of the magnetic field with cluster mass $B \propto M_v^b$, 
with $b \geq 0.6$. 
We also show that the expected occurrence of
radio halos with mass and redshift, and their number counts 
are sensitive to the magnetic field intensity in massive galaxy clusters 
and to the scaling of B with cluster mass. Thus 
future deep surveys of radio halos would provide constraints on B in galaxy clusters.
}

\correspondence{R.Cassano: rcassano@ira.inaf.it}

\maketitle

\section{Introduction}

Giant radio halos (size $\sim 1$ $h_{50}^{-1}$ Mpc, GRHs elsewere) are 
the most spectacular evidence for non-thermal components (relativistic particle and magnetic field) in the intra cluster medium (ICM).
They are diffuse synchrotron radio emission extending on Mpc scale, 
have no obvious connection with the cluster galaxies, but are rather associated 
with the ICM (e.g., Giovannini \& Feretti 2000).
Such a synchrotron radio emission requires a population 
of GeV relativistic electrons (and possibly positrons) and cluster magnetic 
fields on $\mu$G levels. An independent evidence of the existence of
cluster magnetic fields in the ICM is given by the rotation measurement 
studies of radio galaxies located within or behind the clusters (e.g., Govoni \& Feretti 2004). One possibility to explain the presence of GeV electrons 
diffused on Mpc scales is given by the so-called {\it re-acceleration} model.
In this model the radio emitting electrons injected in the 
ICM (by AGN, starbursts, supernovae and/or galactic winds, and hadronic collisions) 
are re-accelerated {\it in situ} by some kind of turbulence generated 
in the cluster volume during cluster mergers (e.g., Brunetti et al. 2001, 2004; 
Petrosian 2001; Fujita et al. 2003; Brunetti \& Blasi 2005). 
Observations indicate that the detection rate of
GRHs shows an abrupt increase with increasing the 
X-ray luminosity of the host clusters. In particular about 30-35\% 
of the galaxy clusters with X-ray luminosity larger than $10^{45}$ erg/s  
show diffuse non-thermal radio emission (Giovannini \& Feretti 2002); 
these clusters have also high temperature (kT $>$ 7 keV) and large mass 
($\gtsim\, 2\times$ $10^{15} M_{\odot}$). Furthermore, GRHs 
are always found in merging clusters (e.g., Buote 2001). 
Although the {\it re-acceleration} model seems to reproduce the 
observational features of the diffuse radio emission, a theoretical investigation 
of the statistical properties of the GRHs in galaxy clusters in the framework of
this model has not yet carried out extensively.
Only recently, Cassano \& Brunetti (2005; CB05 elsewhere), 
have calculated the expected occurrence of GRHs as a function of the
cluster mass and dynamical status in the framework of the re-acceleration
model. CB05 follow cluster formation using the Press \& Schechter (1974; 
PS hereafter) formalism and assume that a fraction, $\eta_t$, of the 
$PdV$ work done by the merging subclusters in going through the main one 
is injected in the form of fast magnetosonic (MS) waves which accelerate 
relativistic electrons in the ICM. They show that GRHs are 
{\it naturally} expected only in the more massive clusters
with an expected occurrence (at $z\ltsim\, 0.2$) which can be 
reconciled with the observed one under viable assumptions ($\eta_t\,\simeq\,0.24-0.34\,$). 
More recently, Cassano, Brunetti, Setti (2006, CBS06 elsewhere)
extended this investigation and calculated the expected evolution
with cosmic time of the fraction of galaxy cluster with GRHs, the expected
luminosity function of GRHs and their number counts. In CBS06 it was assumed
a scaling of the rms magnetic field with cluster mass $B \propto M^b$   
($b<2$) and it was shown that the available radio-X correlations
can be accounted for by assuming $b\ge0.5-0.6$.
A scaling of the magnetic field with cluster mass is indeed expected
by numerical MHD simulations (e.g., Dolag et al. 2004) where the
seed field is amplified by the effect of shear flows 
driven by the cluster formation process.
These simulations show that the resulting magnetic field strengths 
depend on the final cluster mass and/or temperature as $B\propto T^a$ 
with $a\sim2$ (Dolag et al. 2002).
Here we focus our attention on the possibility to make constraints
on the value of the magnetic field strength in galaxy clusters 
by comparing the observed and the predicted statistical properties of GRHs. 
In addition, we show how expectations for future observations will
be powerful tools to test this re-acceleration model and also to provide
additional constraints on B in the ICM. We focus our attention on 
GRHs only (size $\sim 1$ $h_{50}^{-1}$ Mpc,
GRH elsewhere).
The adopted cosmology is: $\Lambda$CDM ($H_{o}=70$ Km $s^{-1}$ $Mpc^{-1}$, $\Omega_{o,m}=0.3$,
$\Omega_{\Lambda}=0.7$, $\sigma_8=0.9$).

\section{X-ray--radio correlations and constraining B}

It is well known that the radio power of GRHs scales with cluster
mass, X-ray luminosity and temperature (e.g., Feretti 2003). Using a sample 
of 17 galaxy clusters CBS06 found a correlation between the radio power 
at 1.4 GHz and the cluster virial mass $P_{1.4}\propto 
M_v^{\alpha_M}$, with $\alpha_M=2.9\pm 0.4$. 

\begin{figure}
\resizebox{\hsize}{!}{\includegraphics{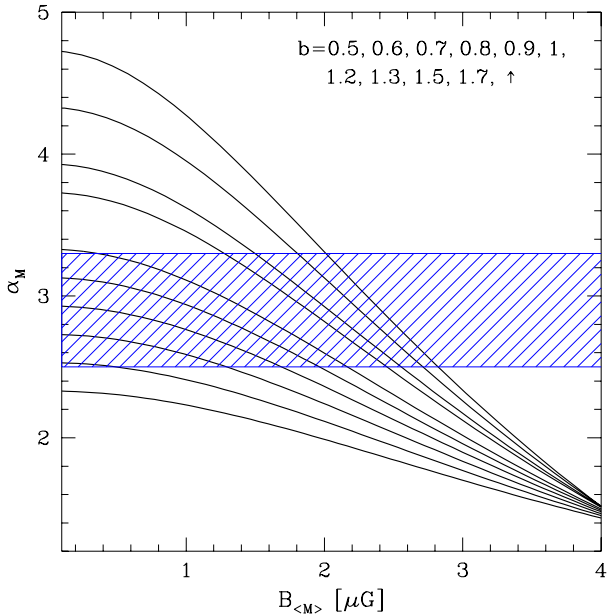}}
\caption[]{Expected slope of the $P_{1.4}-M_v$ correlation as a function of the magnetic 
field intensity, for $<M>=1.6\times\,10^{15}\,M_{\odot}$ and $\Gamma\simeq2/3$. 
The calculations are obtained for $b=0.5-1.7$ (from bottom to top, see panel).
The shadowed region indicate the 1 $\sigma$ range of the observed slope $\alpha_M$.}
\label{expslope}
\end{figure}

\begin{figure}
\resizebox{\hsize}{!}{\includegraphics{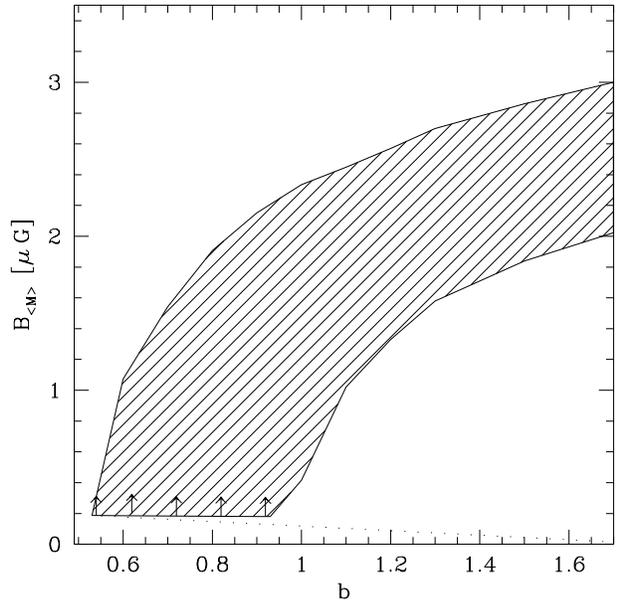}}
\caption[]{Region in the plane ($B_{<M>}$, b) allowed from the observed 
$P_R-M_v$ correlation. Vertical arrows indicate the lower limits 
on $B_{<M>}$ inferred from IC argouments (see text).}  
\label{regions}
\end{figure}

By assuming that electrons are re-accelerated in the ICM, 
the aim of this Section is to constrain the magnetic field 
strength in galaxy clusters and its dependence on cluster 
mass by comparing this correlation with that 
expected by the electron re-acceleration model.
CB05 derived an expected trend between 
the bolometric radio power, $P_R$, and the virial cluster's mass 
in the framework of the particle acceleration model.
Assuming a scaling of B with the cluster mass of the form 
$B=B_{<M>}(M/<M>)^{b}$ (with $b>0$ and $B_{<M>}$ being the rms magnetic
field associate to a given cluster mass, here 
$<M>\,\simeq1.6\times10^{15}M_{\odot}$) one has:

\begin{equation}  
P_{R}\propto \frac{M_v^{2-\Gamma}\,B_{<M>}^2\cdot (M_v/<M>)^{2b}}
{(B_{<M>}^2\cdot(M_v/<M>)^{2\,b}+B_{cmb}^2)^2}
\label{PMpre2}
\end{equation}

where $B_{cmb}=3.2 (1+z)^2 \mu G$ is the equivalent magnetic field 
strength of the CMB and $\Gamma$ is defined by 
$T\propto M^{\Gamma}$; we use $\Gamma\simeq 2/3$ (virial case) or 
$\Gamma\simeq0.56$ (e.g., Nevalainen et al. 2000).
The expected value of the slope of Eq.\ref{PMpre2} depends on $B_{<M>}$ and $b$.
CBS06 show that Eq.~\ref{PMpre2} can be used also for the scaling between
the monochromatic (observed)
radio power at 1.4 GHz and the cluster mass, so that the rms magnetic field
and b can be constrained by matching the slope of Eq.~\ref{PMpre2} with the observed
one, $\alpha_M=2.9\pm0.4$. In Fig.~\ref{expslope} we report 
the expected $\alpha_M$ as a function
of $B_{<M>}$ for $\Gamma\simeq 2/3$ for different values of $b$ 
($b=0.5$ to $1.7$, see caption) together with the 
region spanned by the observed 1 $\sigma$ range.
It is clear that the observed values of the slope cannot be reconciled
with the expected one for $b<0.5-0.6$ and that larger values of
$B_{<M>}$ are required by increasing b.

In Fig.~\ref{regions} we report the allowed region of parameters
in the plane ($B_{<M>},b$) (shadowed area) which is obtained by 
requiring that the expected slope of Eq.~\ref{PMpre2} is consistent 
with the observed one at 1$\sigma$ level.
An additional lower limit on $B_{<M>}$ (vertical arrow in Fig.\ref{regions})
can be inferred in order to not overproduce, via IC scattering 
of the CMB photons, the fluxes of the hard-X ray excesses 
observed in a few clusters (e.g., Fusco-Femiano et al. 2003).
These values of $B_{<M>}$ should be considered as lower limits 
because the IC emission may come from a more external region with 
respect to the synchrotron emission (e.g., Brunetti et al. 2001; 
Kuo et al. 2003; Colafrancesco et al. 2005) and also because 
additional mechanisms may contribute to the hard-X ray fluxes. 
We use the value of the magnetic field derived for 
the Coma cluster, $B_{IC}\simeq\,0.2\,\mu G$ (Fusco-Femiano et al. 2004) 
and obtain the lower bound of B with cluster mass from the scaling law $B=B_{<M>}(M/<M>)^{b}$.

The resulting ($B_{<M>}$,$b$) region spans a wide range of values 
of B and b. However, given a fixed slope of the B--M scaling 
the constraints on the value of B are relatively tight. We stress again
that this region refers to the case in which electrons are re-accelerated
in the ICM (re-acceleration model). In this case we can identifies 
two allowed regimes: a super-linear scaling ($b>1$) 
with relatively high values of B and a sub-linear scaling ($b<1$)
with lower values of B. 

\section{Magnetic field and the occurrence of GRHs}

In their theoretical approach CB05 and CBS06 
identified GRHs with those objects in their synthetic cluster
population with a synchrotron break frequency $\nu_b\, \:\gtsim\: 200$ MHz
in a region of 1 $Mpc$ $h_{50}^{-1}$ size. The break
frequency can be expressed as a function of the cluster mass and of the rms field
in the emitting volume, $B\propto M^b$ (CBS06):

\begin{equation}
\nu_b\propto M^{2-\Gamma+b}{{B_{<M>}\:<M>^{-b}\:\eta_t^2}\over
{(B_{<M>}^2(\frac{M}{<M>})^{2b}+B_{cmb}^{2})^{2}}}
\label{nub}
\end{equation}

\begin{figure}
\resizebox{\hsize}{!}
{\includegraphics[]{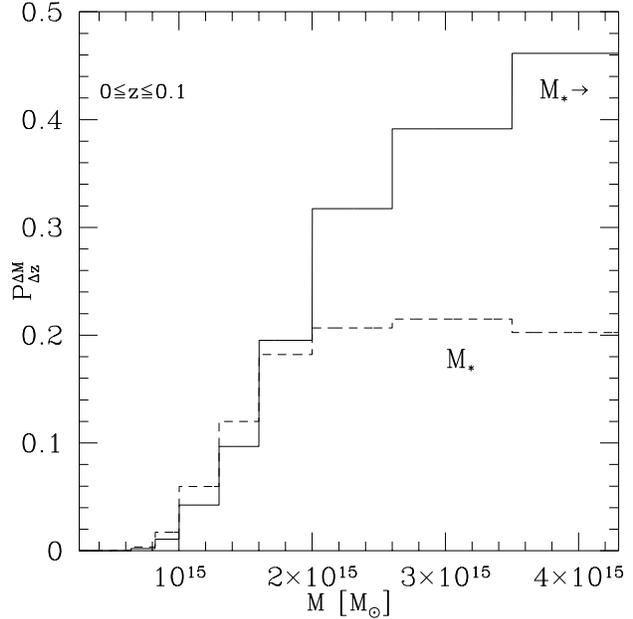}}
\caption{Expected probability to form GRHs as a function of the cluster
mass at redshift $z\le 0.1$. The calculations have been performed
assuming: $b=0.6, B_{<M>}=0.2$ $\mu$G, $\eta_t=0.4$ (solid line) and
 $b=1.7, B_{<M>}=3$ $\mu$G, $\eta_t=0.18$ (dashed line).} 
\label{PM}
\end{figure}

The assumption that B depends on the cluster mass 
should affect the value and dependence with cluster mass
of the synchrotron break frequency and thus the occurrence of GRHs.
Here we are interested in understanding the effect of assuming
different ($B_{<B>}$,$b$) configurations in the resulting probability 
to form GRHs in the context of the particle acceleration model.
Eq.~\ref{nub} has two different behaviors in the case of 
IC dominance ($B<< B_{cmb}$) and in the case of synchrotron 
dominance ($B > B_{cmb}$) in galaxy clusters. 
Typically in the case of a sub-linear scaling 
($b<1$) it is $B<<B_{cmb}$, and an increase of $B$ does not
significantly affect the synchrotron losses. 
In this case one has: $\nu_b\propto M^{2-\Gamma+b}\,(1+z)^{-8}$ and
the probability to form GRHs in these clusters increases 
with mass ($2-\Gamma+b>0$ always) and decreases with $z$.
On the other hand, in the case of a super-linear scaling ($b>1$) 
the value of the mass for which $B$ becomes equal to $B_{cmb}$, 
$M_{*}$ ({\it equipartition} mass), is within the cluster mass range. 
For $M\,>\,M_*(z)$ it is $\nu_b\propto M^{2-\Gamma-3b}$, 
and the particle energy losses would increase
as the mass becomes larger. In this case the probability to 
form GRHs would decrease with cluster mass and the occurrence of GRHs with z 
is mainly driven by the cosmological evolution of the cluster-merger 
history (which drives the injection of turbulence) rather than by 
the dependence of the IC losses with z. 
An example of this different behavior is reported in 
Fig.~\ref{PM} where we plot the probability 
as a function of the cluster mass at $z\le 0.1$ 
for a sub-linear case (solid line, see caption) and
for a super-linear case (dashed line, see caption). 
In the super-linear case, the peak of the probability
is expected at the {\it equipartition} mass $M\sim M_*$.
From the analysis of the fraction of the clusters hosting 
GRH with cluster mass, it is thus possible to constrain the
value of $M_*$ and thus of ($B_{<M>}$, $b$) in galaxy clusters.


\section{Number counts of GRHs and magnetic field}

\begin{figure}
\resizebox{\hsize}{!}
{\includegraphics[]{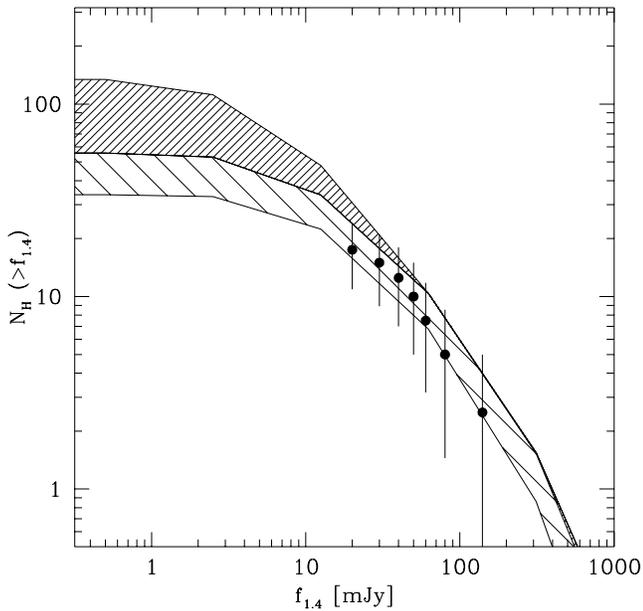}}
\caption{Total number of expected GRHs above a given radio flux at 1.4~GHz from a 
full sky coverage and data points for $z\le 0.2$ (see text).
 Calculations are reported for the superlinear scaling ($b\ge 1$, upper region) 
and for the sublinear scaling ($b\le 1$, lower region).}
\label{conteggi_z02}
\end{figure}

CBS06 derived number counts of GRH in the framework
of the re-acceleration model. The number counts
are obtained by combining the PS mass function of clusters
with the expected radio power--mass scaling (Sect.~2) and with
the probability to form GRHs in a given cluster mass bin.
Since both the probability and slope of the radio power--mass
scaling (Fig.~\ref{expslope}) depend on the scaling $B\propto M^b$
($B_{<M>}$,$b$), the expected number counts should depend on the
strength and scaling of the magnetic field in galaxy clusters.
Fig.~\ref{conteggi_z02} shows the expected number counts in the
cases of a super-linear and a sub-linear scaling of the rms magnetic
field with cluster mass. The two shadowed regions are obtained by combining
the expectations with $b>1$ (upper region) and
$b<1$ (lower region) by assuming the region of ($B_{<M>}$,$b$)
reported in Fig.~\ref{regions}. 
The black points are the number counts of GRHs for $z<0.2$.
They are obtained by making use of the radio data from the analysis
of the 1.4~GHz NVSS radio survey by Giovannini et al. (1999) and by accounting
for the incompleteness of their sky-coverage.
For fluxes larger than 30 mJy the expected number 
counts are close to the counts obtained from the present
observations ($z\le0.2$ GRHs dominate at these fluxes), while at lower 
fluxes present surveys fail in catching the bulk of GRHs.
At these fluxes our expectations are very sensitive to the scaling 
of B with the cluster mass.
In particular, we note that assuming a superlinear scaling of $B$ with cluster
mass, up to $\sim100$ GRHs are expected to be discovered with
future deeper radio surveys. On the other hand,  
the number of these GRHs in sublinear case should be a 
factor of $\sim 2$ smaller.
It is clear that future deep radio surveys will allow to provide constraints 
on B in galaxy clusters.

\section{Conclusions}

In this contribution we have shown that the observed correlations 
between radio and X-ray properties of galaxy clusters, together with
other statistical properties of GRHs (occurrence as a function of mass and
redshift and number counts), may be useful tools to test the particle acceleration
model for the formations of GRHs.
For the first time we have discussed in some details the magnetic field 
intensity and scaling with mass required in the framework of this model.
 
We made use of the statistical calculations developed in CBS06, which are 
based on the re-acceleration of a seed population of relativistic 
electrons by MS waves injected in the ICM during 
merger events (CB05) and which assume a dependence of the rms
magnetic field intensity on cluster mass, $B =B_{<M>} (M/<M>)^b$. 
The main results of this paper are:

{\it i)} A comparison of the expected correlation between the 
radio power and the cluster mass with the observed one allows 
the definition of a permitted region of the parameter space ($B_{<M>}$,b)
in galaxy clusters, with typically small values of B for sublinear 
scaling ($b<1$) and higher values of B for superlinear scaling ($b>1$). 
A lower bound is found at 
$b\sim 0.5-0.6$, while a lower bound $B_{<M>}=0.2\, \mu$G  
is also obtained from IC arguments (Sect.~2). 
A superlinear scaling of $B$ with mass, as expected by MHD simulations (Dolag et al.~2004)
falls within the allowed region. The values of B in the superlinear case are close to (slightly smaller than) those obtained from rotation measurements (e.g., Govoni \& Feretti 2004), which, however, generally sample regions which are even more internally placed than
those spanned by GRHs.

{\it ii)} We have shown that the occurrence of GRHs as a function of mass and redshift
is sensitive to B in galaxy clusters. This probability depends on the 
merging history of clusters and on the relative importance of the 
synchrotron and IC losses,
which indeed depends on the value of the magnetic field.
We show that typically in the case of sublinear scaling of the magnetic field
with cluster mass ($b \sim 0.6-0.9$) the probability to have GRHs 
increases with cluster mass and decreases with redshift, whereas in the case of 
superlinear scalings ($b \sim 1.2-1.7$) more complex behaviors of the probability
with mass and redshift are present.

{\it iii)} We have derived the integral number counts of GRHs at 1.4~GHz
in both the superlinear and sublinear cases.
We find that at higher fluxes ($>30-40$~mJy) the predicted number 
counts are dominated by the $z\le0.2$ clusters discovered in the NVSS.
We estimate that the number of GRHs which would be discovered 
below 30~mJy by future deeper radio surveys (by LOFAR, LWA, and SKA) 
will be up to $\sim 100$ if superlinear scalings of the mass with B hold,
while they will be up to $\sim 50$ in the case of sublinear scaling. 

\section*{Acknowledgments}
This work is partially supported by MIUR under grant PRIN2004.


\begin{thebibliography}{}

\bibitem{} Brunetti, G., Setti, G., Feretti, L., Giovannini, G.: 2001, MNRAS~320, 365
\bibitem{} Brunetti, G., Blasi, P., Cassano, R., Gabici, S.: 2004, MNRAS~350, 1174
\bibitem{} Brunetti, G., Blasi, P.: 2005, MNRAS~363, 1173
\bibitem{} Buote, D.A.: 2001, ApJ~553, 15
\bibitem{} Cassano, R., Brunetti, G.: 2005, MNRAS~357, 1313 (CB05)
\bibitem{} Cassano, R., Brunetti, G., Setti, G.: 2006, submitted to MNRAS (CBS06)
\bibitem{} Colafrancesco, S., Marchegiani, P., Perola, G.C.: 2005, A\&A~443, 1
\bibitem{} Dolag, K., Bartelmann, M., Lesch, H.: 2002, A\&A~387, 383
\bibitem{} Dolag, K., Grasso, D., Springel, V., Tkachev, I.: 2004, JKAS~37, 427
\bibitem{} Feretti, L.: 2003, in: S. Bowyer, C.-Y. Hwang (eds.), 
{\em Matter and Energy in Clusters of Galaxies}, ASP Conf. Series~301, p.~143
\bibitem{} Fusco-Femiano, R., Dal Fiume, D., Orlandini, M., et al.: 2003, in: 
S. Bowyer, C.-Y. Hwang (eds.), {\em Matter and Energy in Clusters of Galaxies},
ASP Conf. Series~301, p.~109
\bibitem{} Fusco-Femiano, R., Orlandini, M., Brunetti, G., et al.: 2004, ApJ~602, 73
\bibitem{} Giovannini, G., Tordi, M., Feretti, L.: 1999, NewA~4, 141
\bibitem{} Giovannini, G., Feretti, L.: 2000, NewA~5, 335
\bibitem{} Govoni, F., Feretti, L.: 2004, Int. J. Mod. Phys. D~13, 1549
\bibitem{} Kuo, P.-H., Hwang, C.-Y., Ip, W.-H.: 2003, ApJ~594, 732
\bibitem{} Nevalainen, J., Markevitch, M., Forman, W.: 2000, ApJ~532, 694
\bibitem{} Petrosian, V.: 2001, ApJ~557, 560
\bibitem{} Press, W.H., Schechter, P.: 1974, ApJ~187, 425

\end{thebibliography}
\end{document}